# Short Communication: The effect of cooling rate and grain size on hydride microstructure in Zircaloy–4

R. Birch, S. Wang, V. Tong, and B. Britton*

Department of Materials, Imperial College London, SW7 2AZ

*Corresponding author: twitter: @bmatb; email – b.britton@imperial.ac.uk

## Abstract

We explore the distribution, morphology and structure of zirconium hydrides formed using different cooling rates through the solid state Zr+[H] → Zr + hydride transus, in fine and blocky alpha Zircaloy-4. We observe that cooling rate and grain size control the phase and distribution of hydrides. The blocky alpha (coarse grain, > 200 µm) Zircaloy-4, has a smaller grain boundary area to grain volume ratio and this significantly affects nucleation and growth of hydrides as compared to fine grain size (~11 µm) material.

## Main Body

Zirconium alloys are used in the nuclear industry as fuel cladding, as it has a good strength to neutron absorption cross section ratio and reasonable corrosion resistance. One concerns when using zirconium alloys in high temperature water reactors is corrosion, as during service it can react with high temperature water to generate an oxide scale and pick up hydrogen [1]. For service conditions (~350°C) this hydrogen may exist in solution where it is highly mobile [2,3]. The hydrogen travels from hot to cold regions and from low to high hydrostatic stress. As the solubility of hydrogen in zirconium has a steep incline against temperature, and so excess hydrogen or changes in temperature may result in the precipitation of hydrides.

The precipitation and growth of these hydrides is both microstructurally sensitive [4–7], and can be related to stress [8,9]. Hydride nucleation and growth is important for delayed hydride cracking (DHC) which is a critical safety case for many water reactors [3]. Temperature gradients often exist within components and thermal gradients may be radial for a fuel tube (internal to external) and this may lead to the formation of a hydride rim at the outer edge. Concentrations of hydrogen within this rim may be three times larger than the bulk concentration [10]. However, knowledge of the hydride rim is limited, including the phase, crystallographic character and mechanical properties. Furthermore, if a supersaturated hydride rim forms, redissolution can occur in thermal conditioning (i.e. heating and cooling to aid dissolution, diffusion and reprecipitation) and the hydrogen may rapidly diffuse to other parts.

The properties of the hydride packets (i.e. clusters of hydride precipitate plates) depend on the microstructure and cooling rate applied to the material. Within zirconium and Zircaloy-4, δ phase (FCC) hydride tends to precipitate at relatively low cooling rates (< 10 °C/min) [11], whereas at cooling rates of > 10 °C/min the γ (FCT) phase hydride is more predominant [12]. For a cooling rate of 1 °C/min, synchrotron XRD based peak deconvolution also implied that δ phase (FCC) hydride could be present, though in the Courty et al. study no other hydride phases were discussed [13]. In Blackmur et al. [3] controversy regarding the nucleation route for δ phase is discussed, but again only the δ phase was analysed in their synchrotron XRD based analysis of hydride solubility and phase fraction using a Rietveld refinement with *a priori* supposition about the presence of α-Zr and δ-hydride. Furthermore, a hydrogen-rich ε phase hydride also exists [14].

This motivates our study of the nucleation, growth and impact of hydrides, especially now that we can generate very large grain 'blocky alpha' [15] where we can vary the area fraction of easy hydride nucleation sites (i.e. grain boundaries) and reduce the interference from one nucleation event to the others by increasing the physical distances between their sites. Ultimately, we hope this promotes our understanding of hydride formation behaviour and mechanism can be extracted from the post-hydriding microstructural investigations.

Zircaloy-4 is a dilute zirconium alloy with a nominal composition of Zr-1.5Sn-0.2Fe-0.1Cr (wt%) [16]. Sn atoms reside as interstitial solute at the interstitial sites of the hexagonal close packed (HCP) α-Zr, while small number of Zr(Fe,Cr)$_2$ secondary phase particles (SPP) of around 100 nm in diameter are also present [17].

Samples of 4 x 3 x 3 mm$^3$ Zircaloy-4 were cut from a rolled and recrystallized sheet, which had an average grain size of ~11 μm and a split basal texture. Half the samples were heat treated at 800 °C for 14 days to promote blocky alpha grain (>200 μm in average size) growth. Representative EBSD crystal orientation maps are shown in Figure 1, for both as-received and large grain materials.

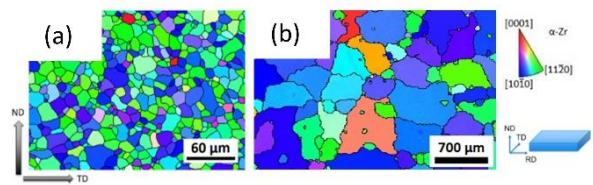

*Figure 1 EBSD crystal orientation maps (IPF-RD, RD direction points outwards of the paper) of as-received (before annealing, left) and large grained (after annealing, right) Zircaloy-4*

Electrolytic hydrogen charging was carried out on each piece of sample individually at 65 °C in 1.5 wt. % sulfuric acid for 24 hours, with a fixed current density of 2 kA/m$^2$ using a Pt mesh counter electrode (see Figure 2a for a schematic of the charging apparatus). This forms a thick hydride rim (Figure 2b). One rim sample was retained for further study (Figure 2d-e, and g).

The remaining samples were annealed in a quartz tube (evacuated and backfilled with Ar) at 400 °C (i.e. super transus for this concentration) for 7 hours to dissolve and homogenise the hydrogen concentration. Samples were returned to room temperature using five different cooling rates (see Table 1) between water quench (i.e. very fast) and a very slow furnace cool (0.05 °C/min).

Estimation of hydrogen concentration was performed with differential scanning calorimetry (DSC, Mettler Toledo DSC1) after metallographic characterisation (X-ray diffraction, XRD; optical microscopy, OM; and electron backscatter diffraction, EBSD). DSC was used to measure the temperature of terminal solid solubility on dissolution (TSSd) upon heating and temperature of terminal solid solubility on precipitation (TSSp) upon cooling of each sample. We used 5 °C/min heating and cooling rates with a peak temperature of 500 °C. A large grained sample which was not hydrided was analysed with DSC as a reference, and no TSSd or TSSp peaks were observed.

DSC results are shown in Table 1 (and the curves are shown in the supplementary data), where the relationship between hydrogen concentration and TSSd in fully recrystallized Zircaloy-4 from [18] is used. TSSp is included for completeness. Variation of the apparent hydrogen content between samples is likely to arise due to slight changes in charging conditions (e.g. depletion of the electrolyte) and dimensions of the samples.

*Table 1 Cooling rates, and TSSd and hydrogen content of the samples obtained by DSC analysis*

|  |  | Quench | Air cool | 3 °C/min | 1 °C/min | 0.05 °C/min | Unhydrided |
|---|---|---|---|---|---|---|---|
| Fine grain | TSSd (°C) | 339 | 401 | 386 | 342 | 336 | - |
|  | TSSp (°C) | 265 | 327 | 315 | 264 | 254 | - |
|  | H content(ppm-wt) | 116 | 236 | 201 | 121 | 112 | - |
| Large grain | TSSd (°C) | 395 | 376 | 347 | 347 | 361 | N/A |
|  | TSSp (°C) | 317 | 307 | 263 | 263 | 279 | N/A |
|  | H content(ppm-wt) | 222 | 180 | 128 | 128 | 152 | N/A |

Ө-2Ө XRD analysis of the rim sample was carried out on the PANalytical X'pert pro MPD machine (hydride rim sample), where the peaks were compared with zirconium hydride phases available in the x'pert highscore plus database. This diffraction pattern fitted best with ε phase (Figure 2c) which correlates with work by Daum et al. [14] and this was supported by cross correlation based EBSD phase discrimination using template matching of a diffraction pattern captured at 30 kV [19], where a higher correlation peak was observed when comparing with a dynamical based simulation of the ε phase (Figure 2d). Cross correlation was performed using patterns generated in DynamicS (Bruker nano GmbH) and reprojected in Matlab.

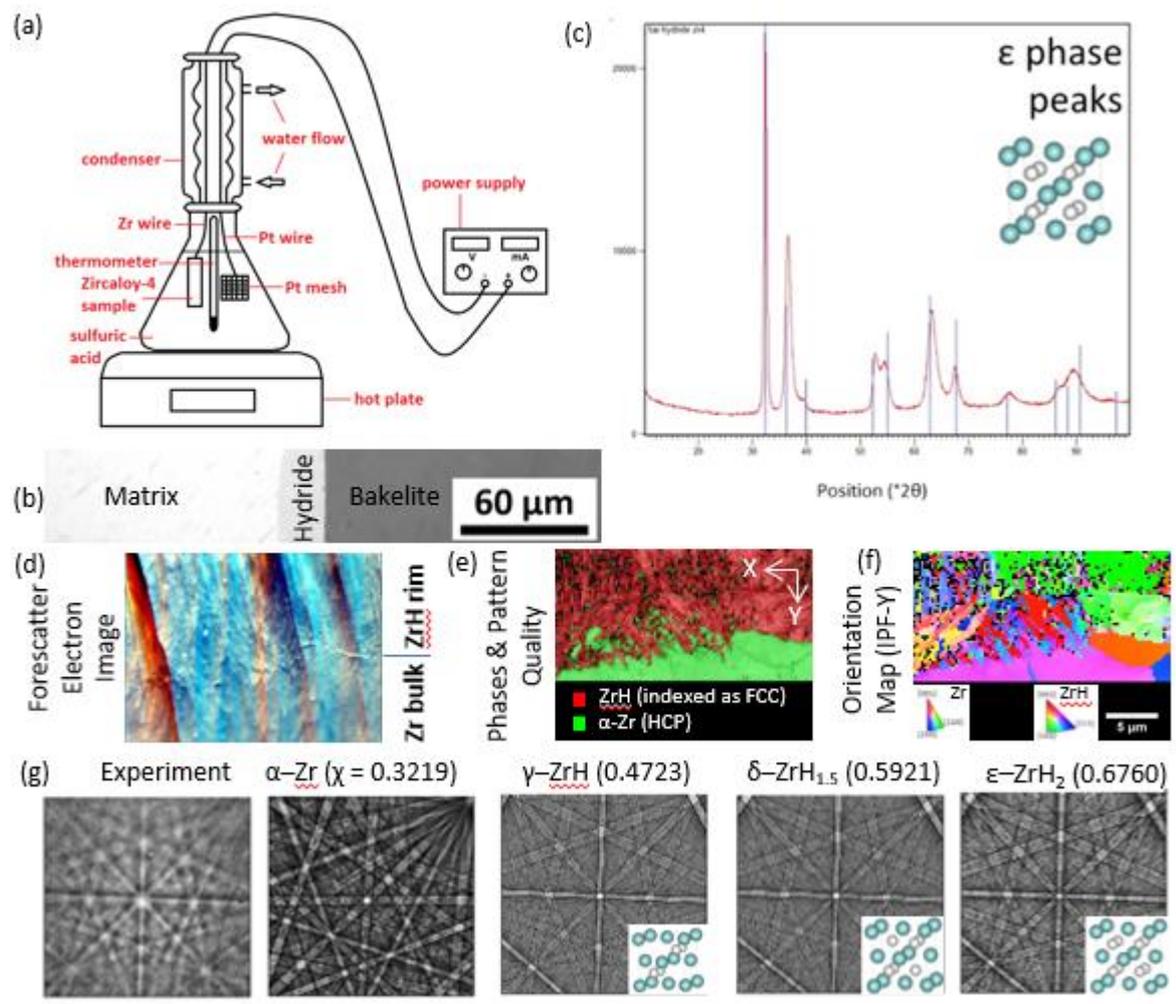

*Figure 2 (a) Schematic of the hydriding rig, (b) hydride rim formed at the sample surface after electrolytic hydriding, (c) XRD pattern for the hydride rim sample, which matches best with epsilon phase (right), (d) Forescatter electron image of the hydride rim, (e) Phase map of the hydride rim overlaid onto the pattern quality map, (f) Crystal orientation map of the hydride rim and (g) comparison of top matches from each hydride phase against the experimental hydride rim EBSD pattern. The normalised cross correlation peak is given in brackets and the higher value indicates a better match.*

Normal and polarised light based optical microscopy of the charged, homogenised and redistributed samples for different cooling rates are shown in Figure 3. Polarised light micrographs show the correlation between hydrides and grains of different orientation and normal optical microscopy highlights the locations of the hydrides within the zirconium.

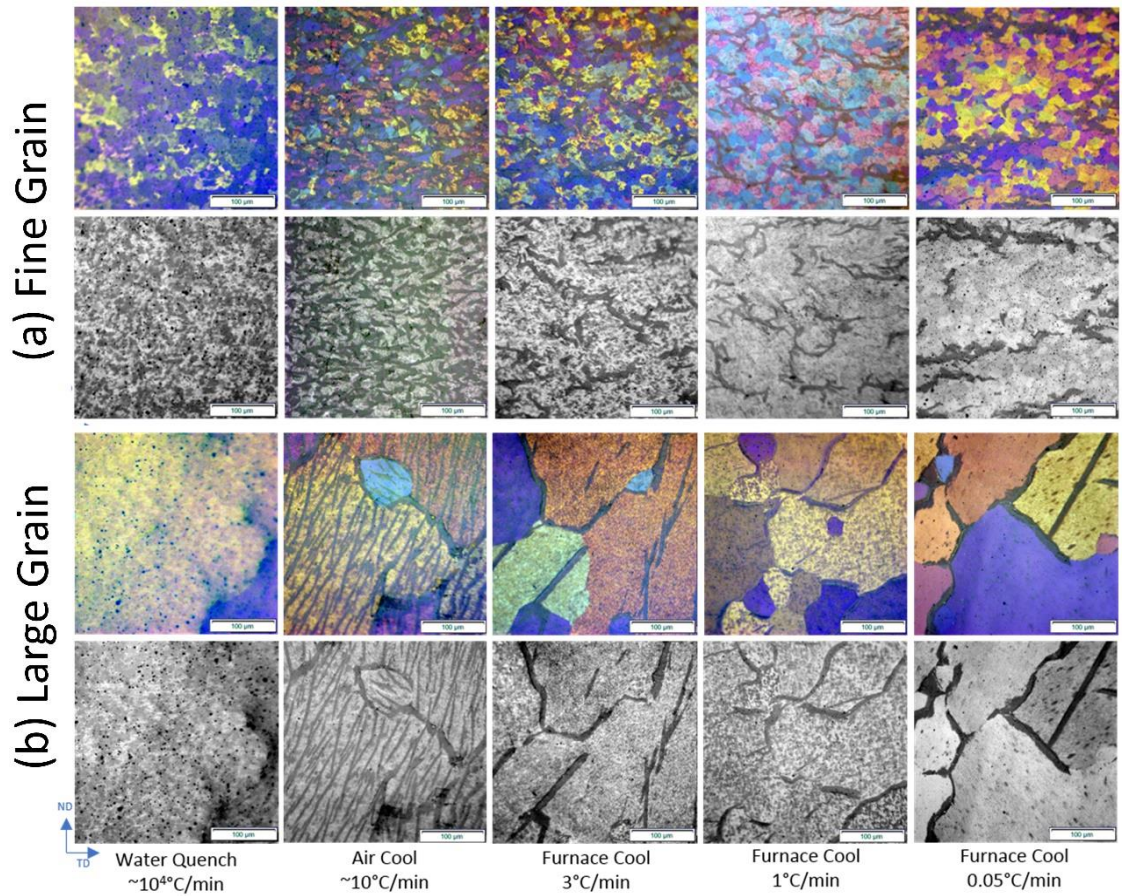

Figure 3: Normal and polarised optimal microscopy of large (blocky) grain and fine grain zircaloy-4 charged between 120-220ppm (see Table 1) and subjected to different cooling rates through the hydride transus temperature.

Water quenching creates a fine distribution of hydrides contained within the matrix for both fine and coarse grained samples. These are well distributed and difficult to resolve directly within the material and it is likely that they distributed homogenously as very fine scale hydrides within the matrix and on the grain boundaries. Scanning electron microscopy (shown in Supplementary Figure 1) indicates that that intragranular hydrides are observed.

Air cooling creates a distinct population of hydrides. In the coarse grained samples these are located both on the grain boundaries and towards the grain interior and they strongly decorate crystal planes. In the fine grained sample thick hydrides are macroscopically aligned along the transverse direction (TD), but this is occluded due to the significant population of intragranular hydrides. Slowing the cooling rates tends to result in clustering of the hydrides, where in the fine grained sample significant 'hydride stringers' are observed and in the coarse grained samples the grain boundary hydrides are readily apparently as dark and coarse features. In the coarse grained samples, EBSD analysis indicates that these hydrides are decorating (near) basal planes of the HCP matrix.

Slower cooling rate develops these trends further and reduces the presence of hydrides not located on the stringer networks (fine grain) or towards the grain boundaries (coarse grain). The grain boundary hydride agglomerates into longer chains or stringers, aligned along TD. For example, at cooling rates of <3°C/min, the hydride distribution changes, with a clear orientation preference developing.

The change in structure and decoration of the hydrides is likely due to a change in the relative undercooling required for nucleation combined with the relative propensity for hydrogen to diffuse towards existing hydride structures and promote growth. This is apparent in the whole sample image and mapping of the hydrides within the slowest cooled (0.05°C/min) blocky alpha microstructure (Figure 4). The EBSD map was aligned with the OM map (Figure 4) to investigate if the alignment of the hydrides could be correlated with grain boundary character (misorientation and surface trace) and unfortunately no correlation of grain boundary character and <c> axis misorientation the hydride decorated boundaries could be observed.

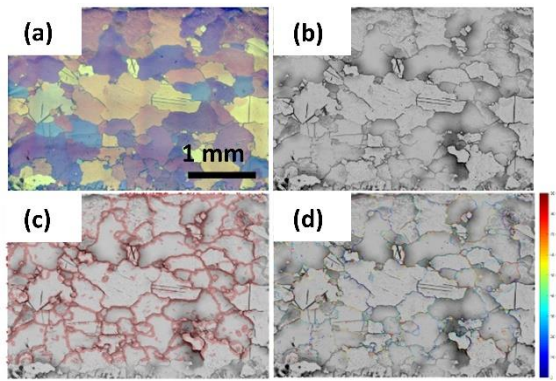

Figure 4: Analysis of slow cooled blockly alpha grains and hydride formation: (a) polarised light micrograph; (b) greyscale micrograph optimised to show hydride locations; (c) hydrides overlaid with EBSD derived grain boundaries, where red grain boundaries have misorientation between the grain boundary trace and the misorientation axis and the thickness correlates with twist (thin) and tilt (thick) character; (d) grain boundaries as highlighted in colour to show <c> axis mismatch.

**Summary**

We observe that the structure and population of hydrides is strongly determined by the cooling rate of when hydrides are precipitated in zirconium. The relative grain size, and thus grain boundary area vs grain interior, can significantly change the hydride population. Importantly in terms of understanding the performance of nuclear fuel, the population of hydrides may influence the failure of the zirconium cladding during delayed hydride cracking and/or long-term storage.

**Acknowledgements**

TBB acknowledges funding from the Royal Academy of Engineering for his research fellowship. TBB and VT acknowledge funding from EPSRC through the HexMat programme grant (EP/K034332/1). Electron microscopy was performed within the Harvey Flower Electron Microscopy Suite and the Quanta was purchased within the Shell AIMS UTC. We would like to thank Alex Foden for assistance with the EBSD pattern matching.

**Data statement**

Data will be released via Zenodo upon article acceptance.

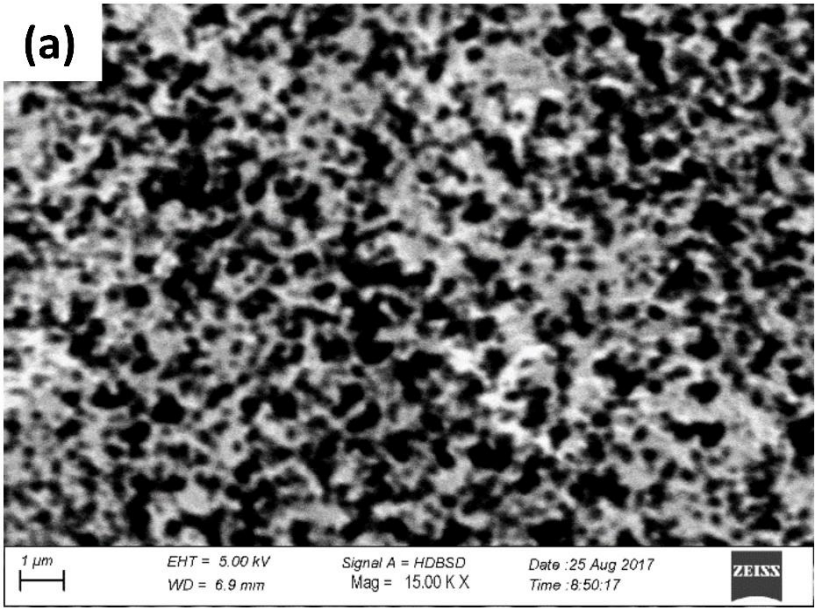

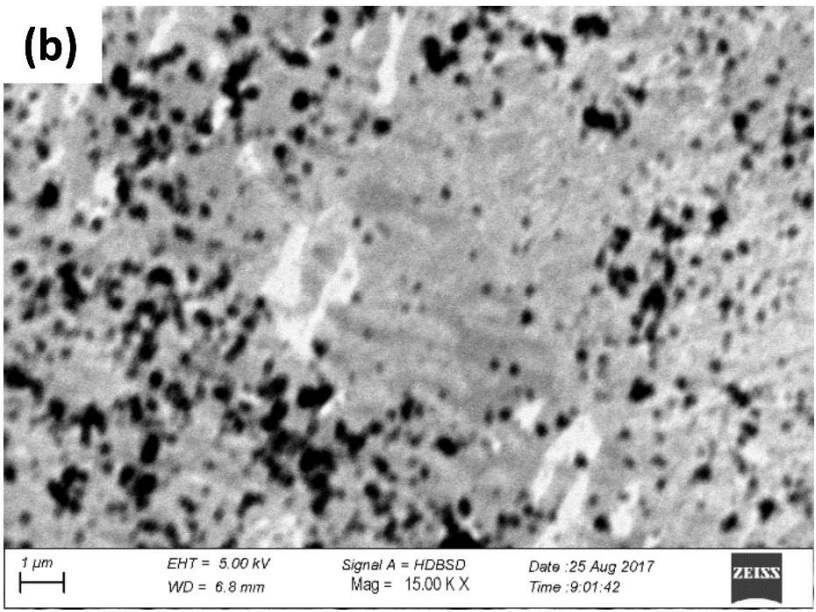

Supplementary Figure 1: Scanning electron micrographs showing the distribution of hydrides in the water quench samples: (a) fine grain; (b) coarse grain. Hydrides appear as black.

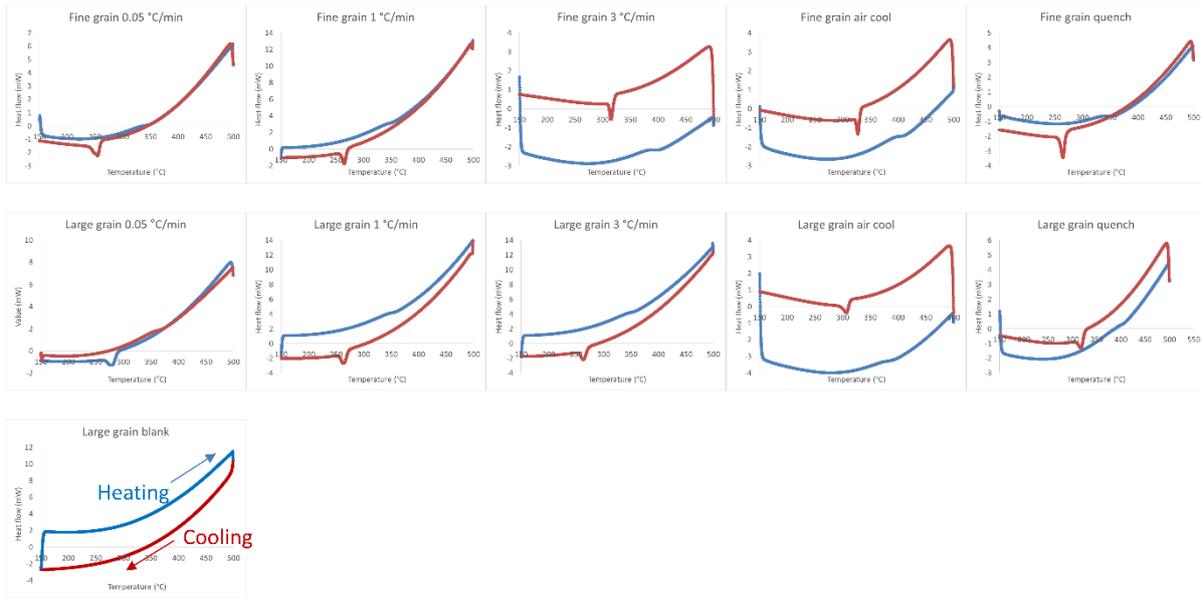

*Supplementary Figure 2: Differential scanning calorimetry curves used to determine TSSp and TSSd for the zirconium hydride samples. TSSp and TSSd are reported in table 1.*